\documentclass[%
 rsi,
 amsmath,amssymb,
 reprint,%
]{revtex4-1}
\usepackage{amsmath}
\usepackage{graphicx}
\usepackage[separate-uncertainty=true]{siunitx}
\usepackage{enumerate}
\usepackage{subfigure}
\usepackage{setspace}
\usepackage{svg}
\usepackage{bm}

\usepackage[utf8]{inputenc}
\usepackage[T1]{fontenc}
\usepackage{mathptmx}
\usepackage{etoolbox}

\begin{document}

\preprint{AIP/123-QED}

\title[Online Charge Measurement for Petawatt Laser-Driven Ion Acceleration]{Online Charge Measurement for Petawatt Laser-Driven Ion Acceleration}
\author{Laura D. Geulig}

 \altaffiliation[Present address: ]{Fakultät für Physik, Ludwig-Maximilians-Universität München, Am Coulombwall 1, 85748 Garching, Germany}
 \affiliation{Accelerator Technology and Applied Physics Division, Lawrence Berkeley National Laboratory, Berkeley, CA 94720, USA}
 \affiliation{Institut für Kernphysik, Technische Universität Darmstadt, Schlossgartenstraße 9, 64289 Darmstadt, Germany}
 
\author{Lieselotte Obst-Huebl}%
 \email{lobsthuebl@lbl.gov}
 \affiliation{Accelerator Technology and Applied Physics Division, Lawrence Berkeley National Laboratory, Berkeley, CA 94720, USA}

\author{ Kei Nakamura}
\affiliation{Accelerator Technology and Applied Physics Division, Lawrence Berkeley National Laboratory, Berkeley, CA 94720, USA}
\author{Jianhui Bin}
\altaffiliation[Present address:]{ State Key Laboratory of High Field Laser Physics and CAS Center for Excellence in Ultra-Intense Laser Science, Shanghai Institute of Optics and Fine Mechanics, Chinese Academy of Sciences, Shanghai 201800, China}
\affiliation{Accelerator Technology and Applied Physics Division, Lawrence Berkeley National Laboratory, Berkeley, CA 94720, USA}

\author{Qing Ji}
\affiliation{Accelerator Technology and Applied Physics Division, Lawrence Berkeley National Laboratory, Berkeley, CA 94720, USA}
\author{Sven Steinke}
\altaffiliation[Present address:]{ Marvel Fusion GmbH, Blumenstrasse 28, 80331 Munich, Germany}
\affiliation{Accelerator Technology and Applied Physics Division, Lawrence Berkeley National Laboratory, Berkeley, CA 94720, USA}
\author{Antoine Snijders}
\affiliation{Biological System and Engineering Division, Lawrence Berkeley National     Laboratory, Berkeley, CA 94720, USA}
\author{Jian-Hua Mao}
\affiliation{Biological System and Engineering Division, Lawrence Berkeley National     Laboratory, Berkeley, CA 94720, USA}
\author{Eleanor A. Blakely}
\affiliation{Biological System and Engineering Division, Lawrence Berkeley National     Laboratory, Berkeley, CA 94720, USA}
\author{Anthony J. Gonsalves}
\affiliation{Accelerator Technology and Applied Physics Division, Lawrence Berkeley National Laboratory, Berkeley, CA 94720, USA}
\author{Stepan~Bulanov}
\affiliation{Accelerator Technology and Applied Physics Division, Lawrence Berkeley National Laboratory, Berkeley, CA 94720, USA}
\author{Jeroen van Tilborg}
\affiliation{Accelerator Technology and Applied Physics Division, Lawrence Berkeley National Laboratory, Berkeley, CA 94720, USA}
\author{Carl B. Schroeder}
\affiliation{Accelerator Technology and Applied Physics Division, Lawrence Berkeley National Laboratory, Berkeley, CA 94720, USA}
\author{Cameron G. R. Geddes}
\affiliation{Accelerator Technology and Applied Physics Division, Lawrence Berkeley National Laboratory, Berkeley, CA 94720, USA}
\author{Eric Esarey}
\affiliation{Accelerator Technology and Applied Physics Division, Lawrence Berkeley National Laboratory, Berkeley, CA 94720, USA}
\author{Markus Roth}
\affiliation{Institut für Kernphysik, Technische Universität Darmstadt, Schlossgartenstraße 9, 64289 Darmstadt, Germany}
\author{Thomas Schenkel}
\affiliation{Accelerator Technology and Applied Physics Division, Lawrence Berkeley National Laboratory, Berkeley, CA 94720, USA}

\date{\today}

\begin{abstract}
Laser-driven ion beams have gained considerable attention for their potential use in multidisciplinary research and technology.
Pre-clinical studies into their radiobiological effectiveness have established the prospect of using laser-driven ion beams for radiotherapy.
In particular, research into the beneficial effects of ultra-high instantaneous dose rates is enabled by the high ion bunch charge and uniquely short bunch lengths present for laser-driven ion beams.
Such studies require reliable, online dosimetry methods to monitor the bunch charge for every laser shot to ensure that the prescribed dose is accurately applied to the biological sample.
In this paper we present the first successful use of an Integrating Current Transformer (ICT) for laser-driven ion accelerators. 
This is a non-invasive diagnostic to measure the charge of the accelerated ion bunch. 
It enables online dose measurements in radiobiological experiments and facilitates ion beam tuning, in particular, optimization of the laser ion source and alignment of the proton transport beamline. 
We present the ICT implementation and the correlation with other diagnostics such as radiochromic films, a Thomson parabola spectrometer and a scintillator. 
\end{abstract}

\maketitle

\section{\label{s:intro}Introduction}
Laser-driven (LD) ion sources have gained attention in various fields of study due to their potential to serve as injectors for conventional accelerators \cite{Krushel2000}, as probe beams for radiography \cite{Borghesi2002a}, as drivers for fusion energy research \cite{Roth2001}, and to provide ion beams for radiotherapy \cite{Bulanov2002}. 
Interest in the latter was sparked because of the considerable prospect to improve the therapeutic effectiveness by making use of unique LD ion beam parameters \cite{Macchi2013, Daido2012}.
One advantage comes from the typically broadband energy spectra that can be shaped into a spread-out Bragg Peak to irradiate a 3D volume in a single shot. 
A second advantage lies in the simultaneous acceleration of protons, carbons and heavier ions from a hydrocarbon contamination layer and the target bulk, which opens up different regimes of biological effectiveness. Throughout this manuscript, the term "ions" refers to both protons and heavier ions.
A third advantage is related to the high bunch charge and uniquely short bunch length of $<$ ps at the source, which makes LD ion pulses valuable tools to investigate the FLASH radiotherapy effect, observed in irradiation studies with ultra-high dose rates \cite{Favaudon2014}.
Recently re-visited after decades of anecdotal reports, the FLASH effect describes the beneficial differential effects on tumors versus normal tissues using the delivery of high radiation doses at extremely high dose rates ($>$40 Gy/s with doses $>$10 Gy delivered in $<$100 ms) \cite{Vozenin2019}. 
Laser-driven proton and heavier ion pulses can deliver several orders of magnitude higher instantaneous dose rates (IDR) than typical conventional (radiofrequency) accelerators, potentially further increasing the differential sparing effect on normal tissue and consequently broadening the therapeutic window for radiotherapy.
Access to conventional experimental and medical machines has been rather limited for this type of research \cite{Durante2014} while the steady increase in available compact LD particle sources has already started to open up new experimental options for systematic radiobiological studies \cite{Friedl2021}.
\newline

In an experiment conducted at the Berkeley Lab Laser Accelerator (BELLA) petawatt (PW) laser facility that was described in Ref.~\cite{Bin2022}, biological cells were irradiated with LD protons at an instantaneous dose rate of 10$^7$ Gy/s up to a total absorbed dose of $>$ 30 Gy by accumulating shots at 0.2 Hz repetition rate. After irradiation, the surviving fractions of normal human prostate and prostate tumor cells were compared and it was demonstrated for the first time that LD protons delivered at ultra-high IDR can indeed induce the differential sparing of normal versus tumor cells \textit{in vitro} for total doses $\geq$ 7~Gy~\cite{Bin2022}. \newline

For this type of study, it is necessary to precisely determine the applied proton dose to the biological samples. Currently, no unified reference dosimetry protocol exists for LD ions, which are unique in their ultra-high IDR and broad energy spectra \cite{Enghardt2018}. Substantial development is still required to turn LD ion sources into a reliable beam delivery technology and current LD ions sources suffer from strong shot-to-shot fluctuations (SSF), significantly exceeding the clinically established dose fluctuation standard of 3-5\% \cite{Linz2007}. Therefore, online dose detectors are essential to monitor the applied dose on every shot, so that variations from the total prescribed dose resulting from SSF can be reduced by \textit{in situ} adjustment of the number of shots applied per sample. If applying Bragg peak ions, i.e. ions that are stopped in the sample, dosimetry cannot be performed behind the sample, so these online detectors need to be positioned in the beam path where they can reduce the proton beam quality, particularly at low initial proton energies. \newline 

Innovative dosimetry methods for radiobiological studies with LD ion sources have been developed that use online, minimally invasive, relative dose detectors, e.g. thin transmission ionization chambers, cross-referenced with independent absolute dosimetry methods like radiochromic films (RCFs) \cite{Bin2019a} or Faraday cups \cite{Richter2011}. These have enabled \textit{in situ} dose-controlled LD proton irradiations of biological cell samples at a relative dose uncertainty below 10\% \cite{Zeil2013}.\newline

Here we present an alternative, fully non-invasive method for online charge measurements of LD ion beams in the form of an integrating current transformer (ICT). To that end an ICT (Bergoz Instrumentation), which had been previously used to measure LD electron bunches generated at the BELLA Center \cite{Nakamura2011,Nakamura2016}, was now characterized for the first time during biological cell sample irradiations with few MeV LD protons at the BELLA PW beamline described in \cite{Bin2022}.\newline

An ICT is a passive current transformer designed to measure the charge of short bunches of accelerated particles with a high accuracy without significant losses \cite{Unser1989}. Charged particle bunches passing through the aperture of the ICT coil induce a current which is temporarily stored in a coaxial capacitor and then delivered to a \SI{50}{\ohm} load, enabling signal readout with an oscilloscope. While the signal trace does not reproduce the incoming temporal ion pulse structure, the bunch charge can be measured by integrating over the signal trace. The ICT is sensitive to the polarity of the charge and only detects charged particles, so the measurement is undisturbed by potential X-ray sources. \newline

In addition to using the ICT as an online diagnostic for the dose applied to the cell samples, we also characterized it as a reliable online diagnostic for ion beam tuning. 
Key proton beam parameters on the cell samples needed to be optimized on a daily basis to ensure optimal beam performance and consistent irradiation conditions from day to day.
While this would usually imply time-consuming changes to the cell irradiation setup to implement diagnostics that could measure these proton beam parameters, we demonstrated that the ICT could be used instead without changes to its configuration from the usual cell irradiation on the beamline. 
This significantly shortened the daily beam tuning phase and overall improved the cell irradiation conditions.

Section \ref{sec:setup} introduces the experimental setup and installation of the ICT for LD ion measurements at the BELLA PW. In section \ref{sec:dosimetry} we describe the characterization of the ICT as an online charge diagnostic for radiobiological studies. Section \ref{s:beam tuning} shows that the ICT can be a versatile tool for ion beam tuning and alignment.

\section{Experimental Setup} \label{sec:setup}
\begin{figure}[htbp]
	\centering
    \includegraphics{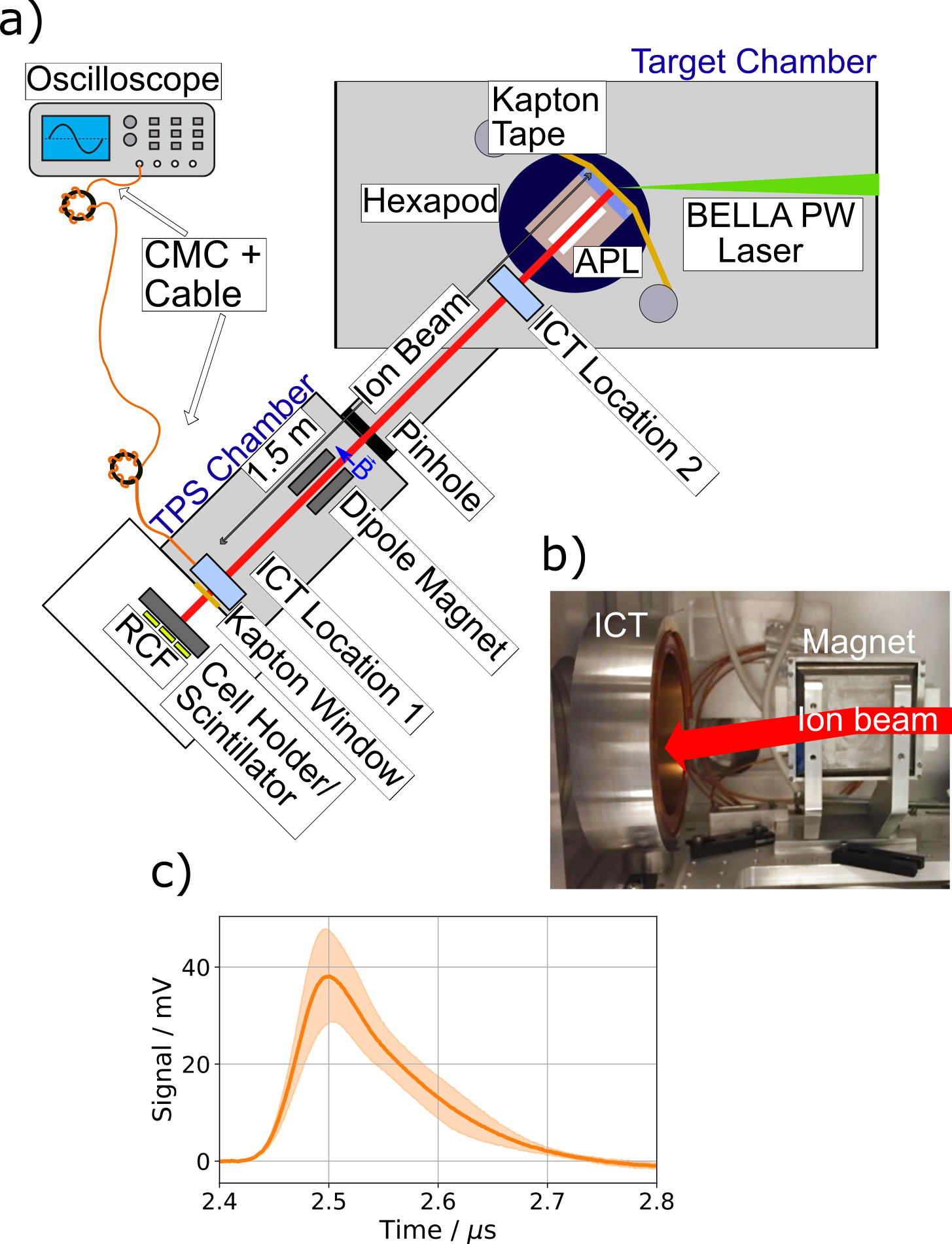}
	\caption{
	(a) shows the schematic setup for laser-driven ion acceleration, ion transport and proton irradiation of biological cell samples. During cell irradiations, the ICT was located in the Thomson parabola spectrometer (TPS) chamber before the beam exited the vacuum system, indicated by "ICT Location 1". In a separate set of measurements, the ICT was positioned at "ICT Location 2" in the target chamber at 30 cm behind the target without the active plasma lens (APL), while a TPS, including a \SI{330}{\mu m} pinhole, a microchannel plate, a phosphor screen and a CCD, was set up in the TPS chamber. (b) depicts the ICT and the permanent dipole magnet in the TPS chamber. The red arrow schematically indicates the ion beam axis. (c) shows the signal trace measured with the ICT and the oscilloscope readout representative of the ion beam charge measurements during cell sample irradiations. The solid line shows an average over 10 shots and the shaded area indicates the standard deviation.}
	\label{fig:radbio}
\end{figure}
\noindent
In the experiment we irradiated biological cells grown in a monolayer on a thin mylar surface within a sample chamber with energetic protons. In order to generate this proton beam, the \SI{40}{J} BELLA PW laser pulse was focused with a F/67 off-axis paraboloid (OAP) on a \SI{13}{\mu m} thick Kapton tape drive target as depicted in Fig. \ref{fig:radbio} (a), which was aligned at an incident angle of \SI{45}{^\circ} with respect to the incoming laser beam. 
The focal spot size on the target was \SI{52}{\mu m} FWHM diameter, yielding a peak intensity of $12 \times 10^{18}~\mathrm{ W/cm^2 }$.

The laser pulses had a tunable pulse length of $t_L$ = \SIrange{35}{>1000}~{fs} full width at half maximum (FWHM) at a central wavelength of $\lambda_L=\SI{815}{nm}$ and a spectral bandwidth of \SI{40}{nm} FWHM generated at the BELLA PW titanium:sapphire laser system with a repetition rate of \SI{1}{Hz} \cite{Nakamura2017}. From the interaction of the laser with the target, ions were accelerated via the target normal sheath acceleration (TNSA) \cite{Wilks2001}. The accelerated ion bunch contained several species with different charge states and  exponential energy spectra with a characteristic cut-off energy. The proton energy cut-off was at \SI{8}{MeV}, consistent with previously published work \cite{Steinke2020}.\newline 

The cell samples were located in air at \SI{1.7}{m} downstream from the point of laser-target interaction.
A portion of the TNSA ion beam was captured and transported downstream by an active plasma lens (APL) on a remote controlled hexapod \cite{VanTilborg2015, Bin2022}. 
The APL consisted of a \SI{33}{mm} long capillary filled with argon gas at a pressure of \SI{5}{Torr} inside the channel of \SI{1}{mm} diameter. 
Applying a discharge current of \SI{90}{\ampere} in the capillary produced a radially symmetric focusing force on the ion beam. 
After the APL the beam passed through a \SI{1}{cm} diameter aperture at 1 m from the TNSA source into the Thomson parabola spectrometer (TPS) chamber, where diagnostics such as a TPS, scintillators, calibrated radiochromic films (RCFs) or the ICT could be installed. 
At the end of the TPS chamber, the beam exited the vacuum system through a \SI{25}{\mu m} thick Kapton window. At this location in the beam path, protons with less than \SI{2}{MeV} and heavier ions were stopped and did not reach the cells or additional diagnostics.\newline

In order to investigate the effect of different doses, groups of cells were irradiated with a varying number of shots at a repetition rate of \SI{0.2}{\hertz}. An RCF sheet (Gafchromic, HD-v2) was placed behind the custom-made cell cartridges so that the exact dose applied to each sample could be determined afterwards. This was possible because the protons were not stopped within the cell cartridges. With this protocol the cells could be regrouped after the irradiation depending on the dose they had received.\newline

During cell irradiations, the ICT (model ICT-122-070-05:1 from Bergoz Instrumentation) was placed in the TPS chamber behind a permanent dipole magnet (refer to "ICT Location 1" in Fig. \ref{fig:radbio}~(b)), which deflected the protons downward from the beam axis. By aligning the cell holders to the deflected protons it was assured that the samples were only irradiated by protons and not by X-rays or electrons. Electrons, originating in the laser-solid interaction and, possibly, the APL discharge, were deflected to such an extent that they did not pass through the 122 mm diameter ICT aperture and did not contribute to the measured signal. 
To mitigate the impact of electro-magnetic pulses (EMP) by the laser-target interaction, double-shielded LMR200 cables in combination with common mode chokes (CMCs, FT-3KL F6045G from Hitachi metals) were used to connect the ICT with the readout oscilloscope (Tektronix DPO 3054). Figure \ref{fig:radbio}~(c) shows a typical ICT signal trace representative of measurements during cell sample irradiations (orange line). This ICT model features a typical rise time of the signal of $\sim$ \SI{36}{ns}.
In order to calculate the charge of the accelerated particles, the trace of the induced current was first smoothed. To account for a constant offset,  the trace recorded prior to the laser-target interaction was averaged and then subtracted. The signal $> \SI{0}{V}$ was then integrated and multiplied with a calibrated conversion factor based on a factory calibration. Before the start of the experimental campaign, this conversion factor was verified using a pulse generator, presenting a well defined charge, connected to a wire going through the ICT's aperture. 
It should be noted that the ICT trace contains all ion species originating in the TNSA source and reaching the ICT aperture by means of the transport beamline.
\newline

In a separate set of measurements, the ICT was located inside the target chamber at 30 cm behind the TNSA source with no APL in front.
This allowed for simultaneous measurements of the ion spectra with a TPS, which are described in Section \ref{s:beam tuning}.

\section{Online Charge Diagnostic for Radiobiological Studies}
\label{sec:dosimetry}
\begin{figure}[htbp]
    \centering
    \includegraphics[width=0.45\textwidth]{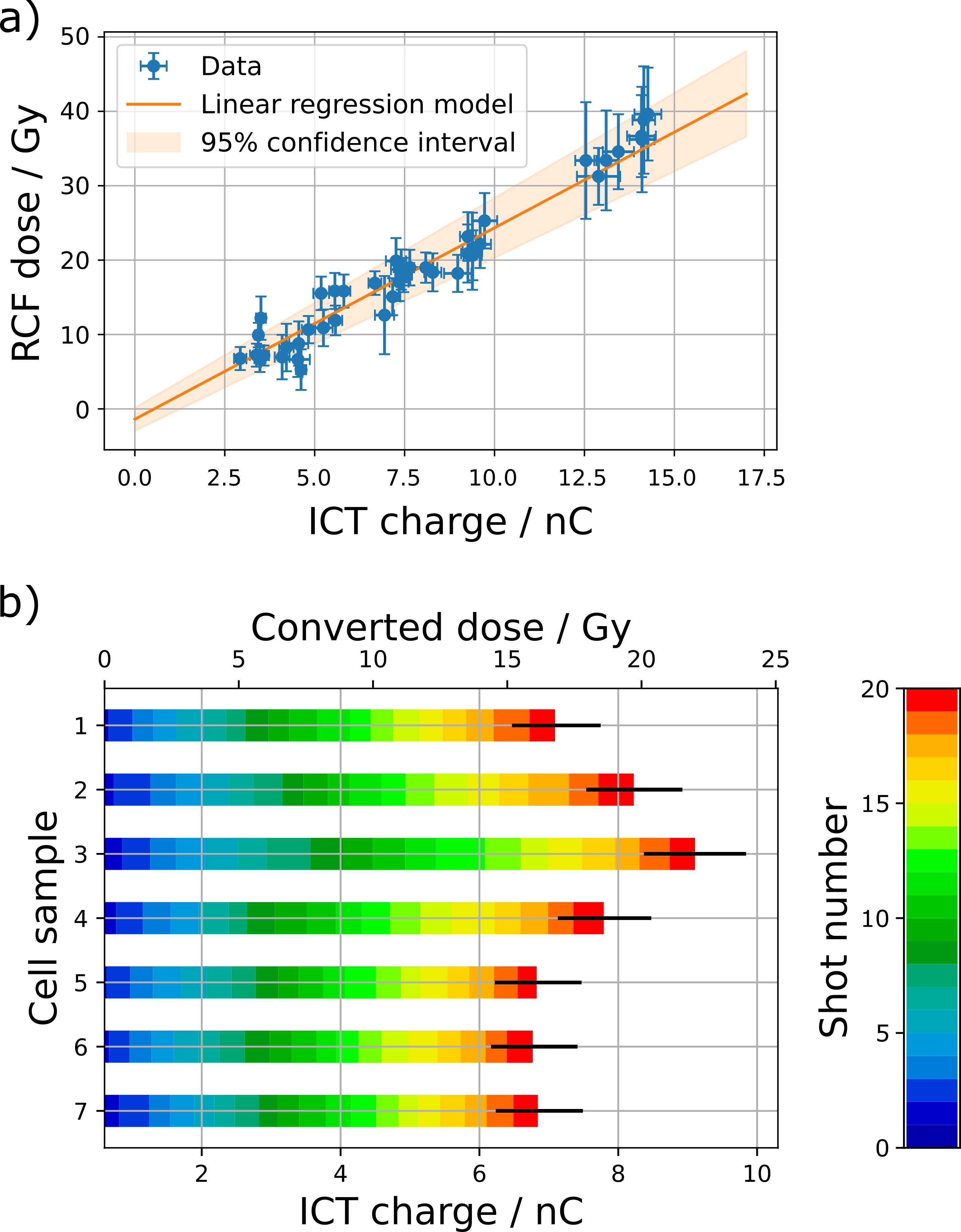}
    \caption{Single radiochromic films behind the cell samples were used to record the dose applied to each cell sample. The ICT charge was integrated and summed for all shots applied to each cell sample, yielding the correlation curve in (a). This correlation makes it possible to use the ICT as a device for online dosimetry. A linear model fit function was applied to allow conversion of ICT measured charge to RCF measured dose on the cell samples. The shaded region indicates the $95\%$ confidence interval of the fit. Error bars represent the standard deviation from averaging over 10 to 30 laser shots.
     (b) shows the measured charge for 20 consecutive shots for seven cell samples, each colored box representing a single shot, with the converted dose on the upper abscissa. Error bars result from the fit uncertainty in (a). This plot displays the influence of shot-to-shot fluctuations on the total dose delivered to different cell samples, varying from \SIrange{16}{22}{Gy}.}
    \label{fig:ICT vs. RCF}
\end{figure}

As mentioned above, single sheets of RCF were placed behind the cell samples outside vacuum to monitor the dose delivered to the cell samples.
Only protons with energies above \SI{2}{MeV} and no heavier ions contributed to the RCF measured dose, while they were detected by the ICT. 
By assuming a stable energy spectrum of the accelerated ion bunch from shot to shot and the fact that the heavy ion acceleration performance typically correlates with the proton acceleration performance in TNSA \cite{Schreiber2006}, a certain charge measured with the ICT is indicative of a specific proton dose on the cells. 
Hence, by comparing the proton dose measured with the RCFs to the charge measured with the ICT, a linear correlation could be established as shown in Fig. \ref{fig:ICT vs. RCF} (a). 
A single data point corresponds to the dose measured by the  RCF  behind the cell sample for the irradiation of that particular sample, accumulated by performing one set of laser shots, versus summed ICT charge measurements from that same set of shots. 
We applied an orthogonal distance regression fit with the linear model function $D(Q)=a\cdot Q +D_0$ with the dose applied to the cells in Gy, $D$, and the charge measured with the ICT in nC, $Q$. The fit parameters were $a=(2.57 \pm 0.12)\ \mathrm{Gy/nC}$ and $D_0=(-1.38 \pm 0.78)\ \mathrm{Gy}$. 
Note that the linear relationship between dose and charge is only verified for the dose range of interest for the cell irradiations in this experiment, i.e. for a charge ranging from \SIrange{2.5}{15}{nC} and needs to be repeated for a different range. 
Moreover, the accuracy of this correlation measurement will be  improved in future experiments by using RCFs better suited for the observed dose range and by this reducing the error of the RCF measurement. 
\newline

The amount of charge generated in each shot, and by this also the delivered dose to the cells, was subject to SSF, as can be seen in Fig. \ref{fig:ICT vs. RCF} (b). 
Here the accumulated charge for seven sets of shots with 20 shots each, to irradiate seven cell samples, is displayed together with the corresponding dose as derived from the correlation model function $D(Q)$ displayed in Fig. \ref{fig:ICT vs. RCF} (a). 
Each colored box represents the charge of a single shot, with the total dose delivered to each cell sample varying from \SIrange{16}{22}{Gy} due to SSF in the ion beam performance. 
This resulted in a relative dose uncertainty of $(\Delta D/D)_{\text{SSF}}=0.17 \pm 0.08$ for irradiated cell samples of the same nominal prescribed dose group.
To reduce the effect of SSF, in this campaign the samples were sorted into groups with similar applied doses based on the RCF-measured dose.
Through sorting, the relative dose uncertainty resulting from SSF between cells of the same nominal prescribed dose group was reduced to $(\Delta D/D)_{\text{SSF}}^{\text{sort}}=0.14 \pm 0.08$.
For future cell irradiation campaigns with the newly established online ICT charge diagnostic, the number of accumulated shots can be adjusted \textit{in situ} to more accurately reach the targeted dose. 
At the current uncertainty associated with the model function for $D(Q)$, $(\Delta D/D)_{\text{SSF}}$ could be lowered to $\sim 0.11$ for total doses $>$\SI{2.5}{Gy} and even to $<0.1$ for doses $>$\SI{10}{Gy}.
It should be pointed out that the relative dose uncertainty resulting from lateral dose variations across the cell samples was $0.19$ \cite{Bin2022} and will also need to be reduced in the future to improve the overall irradiation precision, however, those efforts are beyond the scope of this paper.\newline

To summarize, we found that we can reduce the effect of SSF on the dose precision during cell sample irradiations via the online ICT ion beam charge measurement. This will help us to reach the clinically established dose fluctuation standard of $3 - 5\%$ and, hence, significantly improve future experimental capabilities to study the radiobiological effect of FLASH irradiations at the BELLA PW.

\section{Online Diagnostic for Beam Tuning and Alignment} \label{s:beam tuning}
To ensure best proton beam performance and consistent cell sample irradiation conditions between different days of the campaign, two experimental parameters were optimized on a daily basis. 
First, the laser pulse length was varied to maximize the proton cut-off energy by tuning the separation between the two gratings in the final laser pulse compressor.
Consistent cut-off energies are characteristic for similar particle distributions in the spectral range applied to the cell samples.
Second, the particle number on the \SI{1}{cm} diameter cell samples was optimized by tuning the alignment of the APL with respect to the TNSA source using the motorized hexapod.
It should be noted that in the case of shorter focal length OAP's, the target location along the laser axis is usually another parameter to be optimized on a daily basis.
In our case of the F/67 OAP, the best focal plane only needed to be established once because of the $\sim$\SI{1}{cm} long Rayleigh range and the use of a tape drive target, which reproducibly spooled fresh targets into place \cite{Shaw2016}.\newline

Usually, in order to optimize the laser pulse length or the APL alignment, a TPS in combination with a microchannel plate and a phosphor screen, or a scintillator at the location of the cell samples, respectively, were employed.
This, however, implied significant changes to the cell irradiation setup that were too time-consuming to execute on a daily basis before cell sample irradiations.
Instead, the ICT was established as an online TNSA source optimization and APL beamline tuning diagnostic, by correlating the ICT-measured charge to the proton cut-off energy measured with the TPS and the proton signal at the cell sample location measured with the scintillator.\newline

\begin{figure}[htbp]
    \centering
    \includegraphics[width=0.45\textwidth]{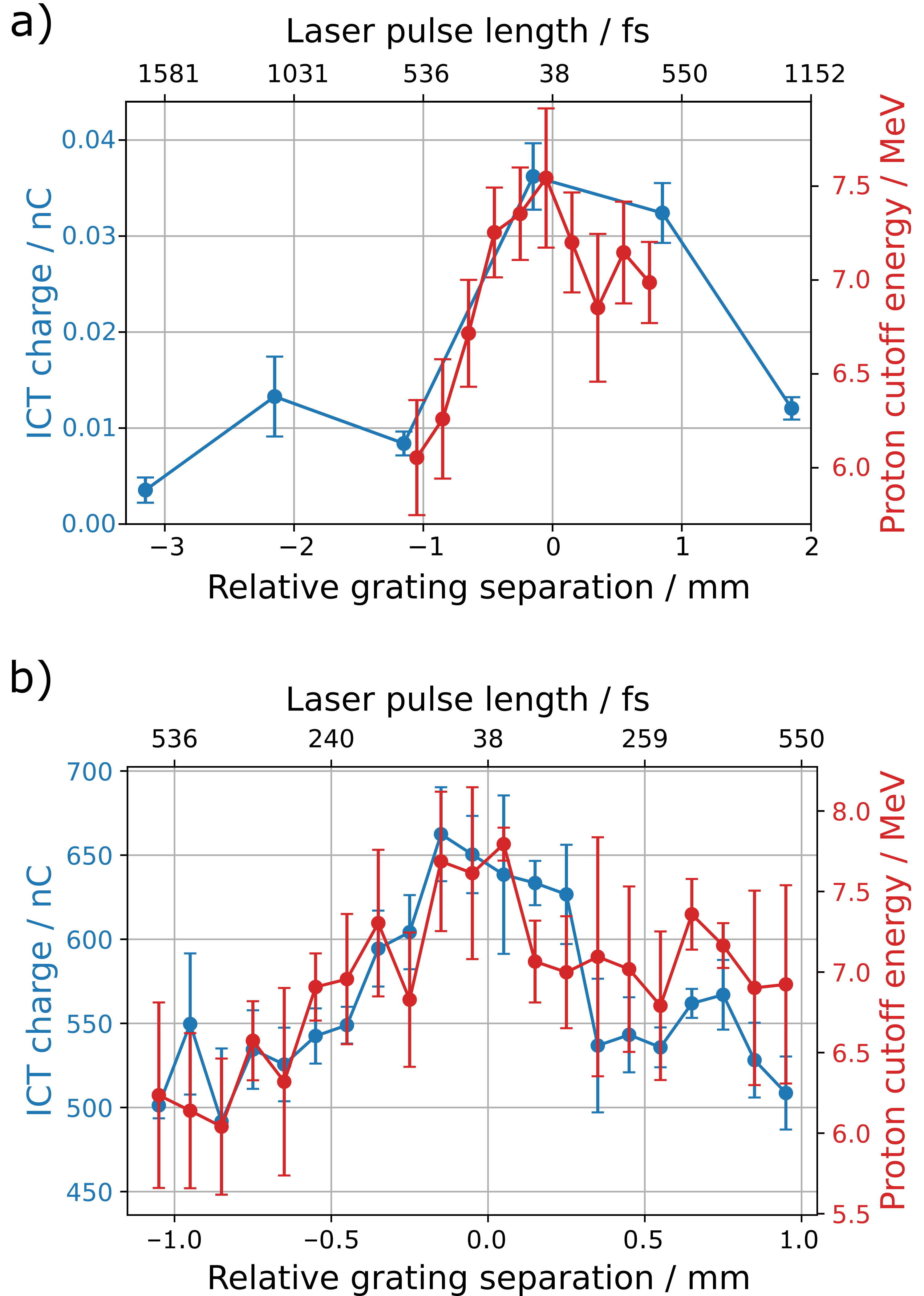}
    \caption{The ICT allows for determining the optimal separation between the two gratings in the compressor of the second CPA stage, which controls the pulse length to optimize for the maximum proton cut-off energy. This was confirmed over a wide range and a non-simultaneous charge measurement with the ICT in the cell irradiation setup and the TPS in (a), as well for a simultaneous measurement with smaller step size in (b). Each data point represents the average of 2 to 6 laser shots, with error bars corresponding to the standard deviation.}
    \label{fig:compressor}
\end{figure}

For laser pulse length scans the proton cut-off energy derived from the analysis of the obtained spectra provides the red trace in Fig. \ref{fig:compressor} (a), showing that the proton cut-off energy is maximized for the optimal compressor setting. 
A separately performed measurement of the beam charge with the ICT over a wide range in blue in Fig. \ref{fig:compressor} (a) shows a maximum charge measured around the same grating position. In this measurement, a \SI{19}{\mu m} thick aluminium foil was placed in front of the ICT, which blocked ion species heavier than protons and protons of kinetic energy lower than \SI{1.2}{MeV}. In order to verify the correlation, a simultaneous measurement of ICT charge, this time without the aluminium foil, and proton cut-off energies was conducted, as shown in Fig. \ref{fig:compressor} (b). For this the 
TPS was mounted in the TPS chamber with a \SI{330}{\mu m} diameter pinhole in front. 
To ensure a sufficient signal detection of the ICT it was placed in front of the pinhole in the target chamber at $\sim \SI{30}{cm}$ behind the target, indicated by "ICT Location 2" in Fig. \ref{fig:radbio} (a), this time without a dipole magnet for electron filtering and without the APL  in front. The same consistency between ICT measured beam charge and TPS measured proton cut-off energy is observed.\newline

\begin{figure}
    \centering
    \includegraphics[width=0.45\textwidth]{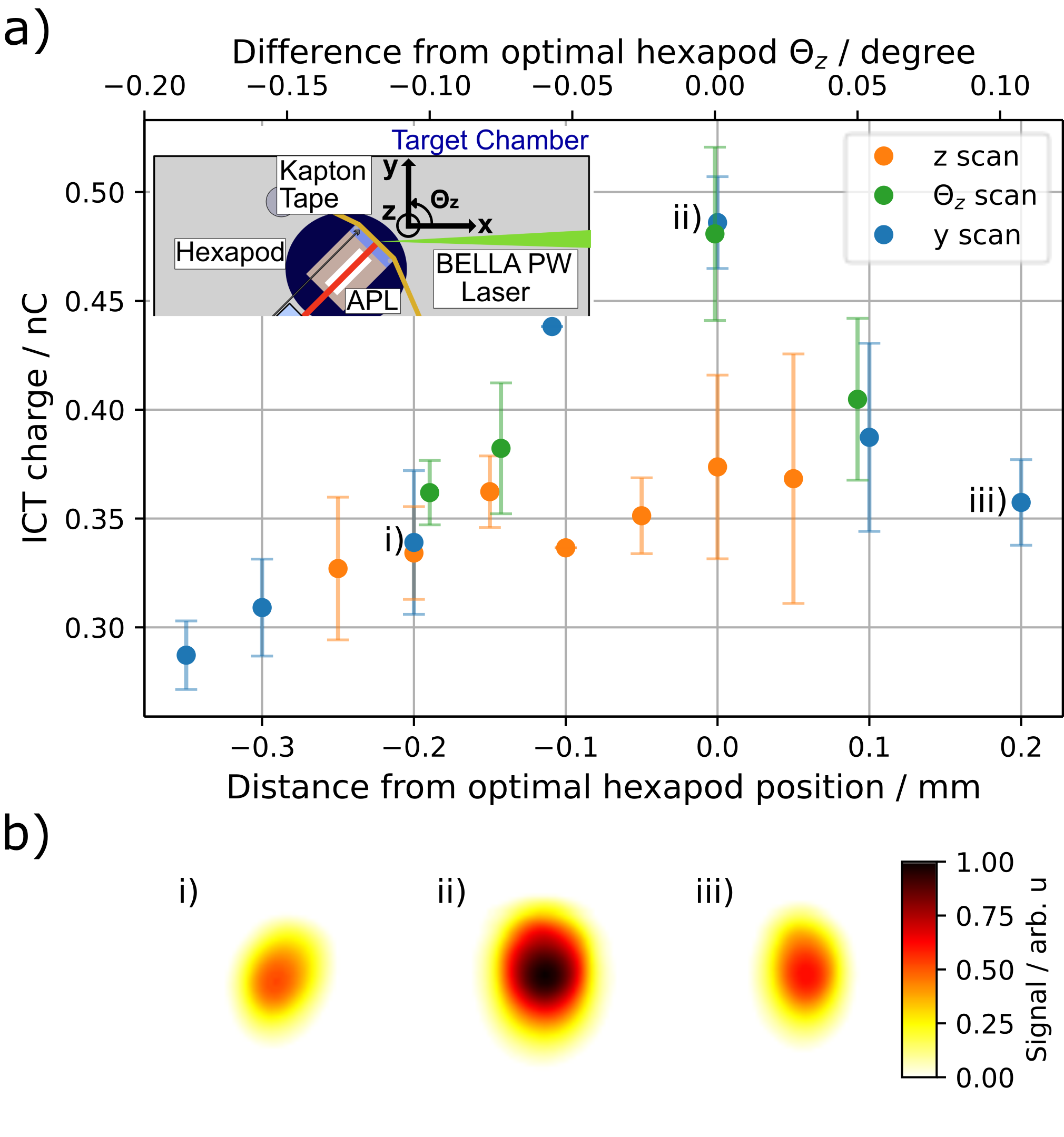}
    \caption{The location of the APL was scanned by moving the remote-controlled hexapod. For this, in (a), first the vertical (z) axis was scanned, the hexapod was then set to the position where the charge was maximized and then the angle around the vertical axis ($\Theta_z$), plotted on the upper abscissa, as well as the transverse (y) axis were scanned for the optimum setting. In parallel to this the proton beam profile was measured with a scintillator at the location of the cell samples that was imaged to a camera. 
    Each data point represents the average of 2 to 8 laser shots, with error bars corresponding to the standard deviation (apart from one data point in the z scan that represents a single shot only).
    (b) Scintillator images taken at the indicated hexapod positions for the scan of the transverse axis in (a) show a clear correlation of the charge measured by the ICT and the signal strength on the scintillator.}
    \label{fig:hex scan}
\end{figure}

A precise alignment of the \SI{1}{mm} diameter aperture of the APL at \SI{13}{mm} behind the target relative to the ion source is crucial to maximize the proton dose on the cell samples. 
Since the exact position of the laser focus and the tape and hence of the ion source could shift between experimental days, the alignment was verified every day. 
For this the light emitted by a scintillator, intercepting the proton beam at the location of the cell samples was maximized, while scanning the position and angle of the APL located on top of a motorized, remote controlled hexapod. 
We found that the optimal position and angle of the APL corresponded to a maximum of signal detected both with the scintillator and the ICT. In Fig. \ref{fig:hex scan} (a) we therefore first scanned the vertical hexapod axis, after the scan moved back to the best vertical position and repeated this procedure for the angle around the vertical axis, plotted on the upper abscissa, and the transverse hexapod positions with respect to the laser beam axis. 
Figure \ref{fig:hex scan}~(b) shows three scintillator images from different transverse APL positions before, at and after the optimum location. \newline

These measurements show that the ICT is an effective diagnostic for tuning of the beam without re-configuring the cell irradiation setup. 
As such, the charge derived from the ICT measurement correlates well with other diagnostics previously used for beam tuning, the use of which, however, require time-consuming changes to the beamline. 
Since the ICT does not affect the ion beam quality, it can remain in the experimental setup and be used whenever it is needed.

\section{Conclusion}
In this paper we presented the first use of an ICT for laser-driven ion acceleration experiments. 
This tool presents a diagnostic for beam charge measurements that, when calibrated, can be used as a device for radiobiological experiments providing information about the delivered dose to a cell sample after every laser shot. 
Its great advantage is that it is non-invasive, thus not degrading the beam quality and that it operates in real time, giving instantaneous feedback compared to the delayed information provided for example by RCFs. 
As such, shot-to-shot fluctuations of the LD ion source performance and the resulting effect on the delivered dose can be accounted for by adjusting the number of shots applied per sample.
In the current setup, the relative dose error can thus be reduced to below 10 \% for total doses $>$\SI{10}{Gy} in future radiobiological studies at the BELLA PW.\newline

We furthermore showed that it is an effective tool for the daily tuning of the ion beam with regard to highest energy and particle number by maximizing the measured charge at the best pulse length and the alignment of the focusing APL.
This enables optimal ion beam parameters that are consistent from day to day, without the need to implement additional diagnostics that require time-consuming changes to the irradiation setup.
As such, the ICT is a versatile diagnostic that will be used as an integral part of the diagnostic suite for future laser-ion acceleration experiments at the BELLA PW.\newline

Future studies with the ICT will establish whether the charge measurement is affected by the very high particle numbers and instantaneous dose rates present for LD ion sources.
So far the linearity of the ICT response was demonstrated for electron bunches of \SI{200}{ps} bunch duration and up to \SI{450}{pC} charge \cite{Nakamura2011}. 
Beyond purely measuring the integrated ion beam charge, the ICT has the potential to be used for time-of-flight measurements of different ion species and unfolding of the spectral information. 
This could be achieved by combining the measurement with simulations of the ICT response by including the ICT circuit architecture. This will be investigated in a future work. 

\begin{acknowledgments}
This work was supported by the Director, Office of Science, Offices of Fusion Energy Science and High Energy Physics of the U.S. Department of Energy under Contract No. DE-AC02-05CH11231. This work was also supported by Laboratory Directed Research and Development (LDRD) funding from Lawrence Berkeley National Laboratory provided by the Director and LaserNetUS (https://www.lasernetus.org/). The authors gratefully acknowledge the technical support from Arturo Magana, Joe Riley, Zac Eisentraut, Mark Kirkpatrick, Tyler Sipla, Jonathan Bradford, Greg Mannino and Nathan Ybarrolaza. The authors would like to thank Sam Barber, Jared De Chant, Zachary Kober, Tobias Ostermayr, Marlene Turner, Csaba Toth and Wim Leemans for their contributions, and Tom Delaviere and Frank Stulle from Bergoz instrumentation for fruitful discussion on integrating current transformers.
\end{acknowledgments}

\section*{Data Availability Statement}

The data that support the findings of this study are openly available on Zenodo, with the digital objective identifier (DOI) 10.5281/zenodo.6419041 \cite{repo}.

\bibliography{bibliography}

\end{document}